\documentclass{aa} 

\usepackage{graphicx}
\usepackage{txfonts}
\usepackage{amsmath}
\usepackage{physics}
\usepackage{xcolor}
\usepackage{float}
\usepackage{hyperref}
\begin{document}

   \title{Multi-threaded prominence oscillations triggered by a coronal shock wave}

   \author{V. Jer\v{c}i\'{c}
          \inst{1}
          \,
          R. Keppens
          \inst{1}
          \and
          Y. Zhou
          \inst{1}
          }

   \institute{\inst{1} Centre for mathematical Plasma-Astrophysics, Celestijnenlaan 200B, 3001 Leuven, KU Leuven, Belgium}

   \date{Received date1 / Accepted date2}

 
  \abstract
   {Understanding the interplay between ubiquitous coronal shock waves and the resulting prominence oscillations is a key factor in improving our knowledge of prominences and the solar corona overall. In particular, prominences are a key element of the solar corona and represent a window into an as yet unexplained processes in the Sun's atmosphere.}
   {To date, most studies on oscillations of prominences have ignored their finer structure and analyzed them strictly as monolithic bodies. In this work, we study the causal relations between a localised energy release and a remote prominence oscillation, where the prominence has a realistic thread-like structure. }
   {In our work, we used an open source magnetohydrodynamic (MHD) code known as MPI-AMRVAC to create a multi-threaded prominence body. In this domain, we introduced an additional energy source from which a shock wave originates, thereby inducing prominence oscillation. We studied two cases with different source amplitudes to analyze its effect on the oscillations. }
   {Our results show that the frequently used pendulum model does not suffice to fully estimate the period of the prominence oscillation, in addition to showing that the influence of the source and the thread-like prominence structure needs to be taken into account. Repeated reflections and transmissions of the initial shock wave occur at the specific locations of multiple high-temperature and high-density gradients in the domain. This includes the left and right transition region (TR) located at the footpoints of the magnetic arcade, as well as the various transition regions between the prominence and the corona (PCTR). This results in numerous interferences of compressional waves propagating within and surrounding the prominence plasma. They contribute to the restoring forces of the oscillation, causing the period to deviate from the expected pendulum model, in addition to leading to differences in attributed damping or even growth in amplitude between the various threads. Along with the global longitudinal motion that result from the shock impact, small-scale transverse oscillations are also evident. Multiple high-frequency oscillations represent the propagation of magnetoacoustic waves. The damping we see is linked to the conversion of energy and its exchange with the surrounding corona. Our simulations demonstrate the exchange of energy between different threads and their different modes of oscillation.}
   {}


   \maketitle
%

\section{Introduction} \label{sec:intro}

   Solar prominences are structures that are often observed in the solar corona, they have a higher density and lower temperature by two orders of magnitude than the background corona. These have been studied and observed for decades \citep{Hyder1966, Kleczek1969, TH1974, Schmitt1995}, but nonetheless, they still represent a mystery in many ways. They are supported against gravity by the Lorentz force resulting from dipped magnetic field lines. A prominence has a magnetic topology of a coronal arcade or a flux rope whose footpoints are rooted in the photosphere. Quiescent prominences, situated outside of active regions, usually last longer than intermediate prominences that are located nearer to active regions, while active prominences are found inside active regions. The average length of prominence is in the range of 30 to 110 Mm, with quiescent ones usually shown to be longer and reaching higher altitudes \citep{Parenti2014}. The quiescent prominences are more stable and their lifetimes range anywhere between several days up to several months, while the active prominences are short-lived \citep[usually shorter than the active region they are related to;  ][]{Labrosse2010}. 

   Even though prominences may appear to be globally stable and static structures, based on multiple observations \citep{Schmieder1991,Engvold1998,Lin2005,Chen2014,Okamoto2016}, we know that they are highly dynamic structures locally. They are made up of multiple threads that have vertical and horizontal flows, with the average lifetime of an individual thread being in the range from a few up to 20\,min \citep{Lin2005,Berger2008}. The average length of prominence threads is between 3.5 and 28 Mm, with an average width of 210 km \citep{Arregui2018}. They are usually observed to be aligned with the prominence magnetic field with a possible small inclination of 20-25$^{\circ}$ to the filament axis \citep[i.e. the filament channel,][]{Lin2005,Luna2018}. There are multiple theories on the formation mechanisms of threads but none have  been  fully verified. Some scenarios explain these threads as resulting from Rayleigh-Taylor instability \citep{Xia2016} or thermal instability \citep{Claes2020}, or otherwise due to random thermal heating at the solar surface \citep{Zhou2020}. 

   Another striking characteristic of prominences is that they are often observed to oscillate due to omnipresent perturbations found in the corona. If we consider the initial velocity amplitudes, according to \cite{Luna2018}, we can group oscillations into large amplitude oscillations (LAOs), with $v \geq$ 10\,km\,s$^{-1}$, and small amplitude oscillations (SAOs), with $v <$ 10\,km\,s$^{-1}$ \citep[for reviews, see also][]{Tripathi2009,Arregui2018}. According to recent statistics carried out by \cite{Luna2018}, based on a six months catalogue of prominence oscillations, both types are common and appear at least every other day on the visible side of the Sun. In most cases, when we talk about LAOs, the whole prominence structure is perturbed, while SAOs affect the prominence only locally. Even though it is generally believed that SAOs are strictly local phenomena \citep{Ning2009,Hillier2013Letter}, \cite{Luna2018} found SAOs affecting the prominence globally. Another of their findings is that both types of oscillations have the same source of perturbation and the mean distribution of their respective periods, while the corresponding standard deviation shows no significant difference. All of these details make it hard to draw a clear distinction between what we call LAOs and SAOs and what tells them apart other than the value of their initial velocity amplitude. 

   An alternative way to categorise prominence oscillations is according to their direction of motion with respect to the background magnetic field. In the case when the motion goes along the magnetic field lines, the oscillations are longitudinal. In the case when the oscillations are perpendicular to the magnetic field lines (either in the vertical or horizontal plane), we talk about (vertical and horizontal, respectively) transverse oscillations. However, because the threads are commonly observed to be inclined to the magnetic field, the coupling of the mentioned modes of oscillation is very frequent. These types of oscillation are damped relatively fast (particularly the longitudinal modes in comparison to the transverse ones). There are multiple theories explaining damping, but the precise mechanism is not fully understood. Most often, it is a combination of different dissipative processes at work, for instance, energy losses due to waves leaking into the surrounding plasma or non-adiabatic effects such as radiative losses and thermal conduction. The most efficient dissipation process also depends on the mode of the oscillation, for example, slow waves are efficiently damped by non-adiabatic effects, while the fast waves remain almost unaffected \citep{Arregui2018}.

   As for the sources of prominence oscillations, according to observations, in most cases these are large-scale shock waves induced by flares, as well as Moreton and extreme-ultraviolet (EUV) waves \citep{Shen2012,Shen2014b} or small flares and jets located near the footpoints of a prominence \citep{Vrsnak2007,Zhang2013}. Interaction between coronal shock waves and prominences is a common phenomenon \citep{Asai2012Letter,Nistico2013,Hillier2013Letter,Mazumder2020}. However, efforts to simulate such events and analyze them more thoroughly have been scarce. Relations between the source properties (e.g. its strength, distance, and position with respect to the prominence) and the resulting oscillations are still fairly unknown. A wave can amplify the already existing oscillations \citep{Zhang2020}, it can disrupt a stable prominence causing it to become activated or eventually erupt \citep{Takahashi2015,Yashiro2020,Devi2021}, or it might not  cause any significant changes at all \citep{Okamoto2004,Shen2014a}. 
   
   A numerical study carried out by \cite{Zhou2018} provided a 3D model of prominence oscillations. The authors embedded a blob of increased density, representing a prominence, into a fully stratified flux rope. They introduced longitudinal and transverse velocity perturbations and studied the resulting motion. Even though the model is in many ways close to reality, it is still missing crucial elements. The type of oscillation is preselected such that the coupling of modes is virtually absent. The prominence is a monolithic body and the resulting motions follow the theoretically expected ones. In reality, the property of prominence being multi-threaded contributes significantly to its observed behaviour. Later, \cite{Zhou2020} carried a 2D, non-adiabatic model of a multi-threaded prominence. However, their goal here was to study the formation mechanisms, so they did not focus on any aspect of prominence oscillations. A recent paper by \cite{Luna2021} has reported on a 2.5D simulation of prominence oscillations caused by a coronal jet. Much as in our approach, which we describe further on in this paper, these authors do not take into account thermal conduction or radiative losses. They find an Alfv{\'{e}}nic front propagating ahead of the jet. Subsequently, multiple reflections and transmissions of the jet fronts through the prominence cause counter-streaming flows. As a result of the jet's impact, the monolithic prominence experienced both transverse and longitudinal oscillations. \cite{Liakh2021} conducted a 2D adiabatic simulation of prominence oscillations to investigate the amplification and attenuation mechanisms of large-amplitude longitudinal oscillations. Additionally, they wanted to find out how grid resolution affects the results of such a study. According to their results, there is a limit to how much numerical damping can influence the simulation results. With a high-enough grid resolution (of about 30\,km), the damping effects are predominantly due to physical reasons. Similarly to our approach, they used a 2D adiabatic model -- but with prominence as a monolithic body, situated in an $x$-$z$ plane (perpendicular to the Sun's surface). The results show an amplification of oscillation in the top parts of the prominence, which they interpreted as a consequence of energy transfer from the bottom (where damping was noticed) to the top regions of the prominence.
   
   In many cases, prominence oscillations end in a violent eruption, ejecting vast amounts of material into space. Such eruptions are known as coronal mass ejections (CMEs), which present a possible threat to Earth. Therefore, studying prominence oscillations becomes highly imperative, as we wish to better understand the cause and effect in eruptive scenarios. To understand the interplay of the fine structure and the corresponding global dynamics of oscillation, we performed a numerical simulation of a localised energy source that drives a shock wave. Unlike any of the previous works studying prominence oscillation, we combined a multi-threaded prominence with a realistic source region. We describe the characteristics of such prominence oscillations in a low-beta regime. We do not select the modes of oscillation in advance but rather allow for coupling of magnetohydrodynamic (MHD) modes in a 2D adiabatic setup. Section \ref{sec:method} gives further details on the numerical setup we used, the geometry of the system, and how we introduce additional energy into the system. In Section \ref{sec:results}, we report on the findings of such simulation and in Section \ref{sec:discussion}, we elaborate on the results. The final section (Section \ref{sec:conclusion}) gives our concluding remarks on the research and findings.

\section{Numerical method} \label{sec:method}

   Quiescent prominences are stable structures situated outside of active regions. They are maintained by the magnetic field anywhere from a few days up to several months. As such, they are the perfect candidate for studying the oscillatory characteristics of prominences with a finer, thread-like structure (also known as fibrils). Even though quiescent prominences are statistically characterised on the basis of a flux rope topology \citep{Ouyang2017}, we used a magnetic arcade model where the fixed magnetic dipped topology is prescribed from the beginning. The arcade configuration will have fully two-dimensional dynamics due to the field-aligned gravity that we plan to investigate here and this model was used previously to study thread formation in a randomly heated arcade \citep{Zhou2020}. In Section \ref{subsec:geometry}, we provide more details on the actual configuration used. We explain how we induce the prominence oscillations in Section \ref{subsec:numerics}, along with the governing equations, implementation, and  details of the algorithm. 

\subsection{Geometrical configuration and induced oscillations} \label{subsec:geometry}

\begin{figure*}
   \centering
   \includegraphics[width=0.95\textwidth]{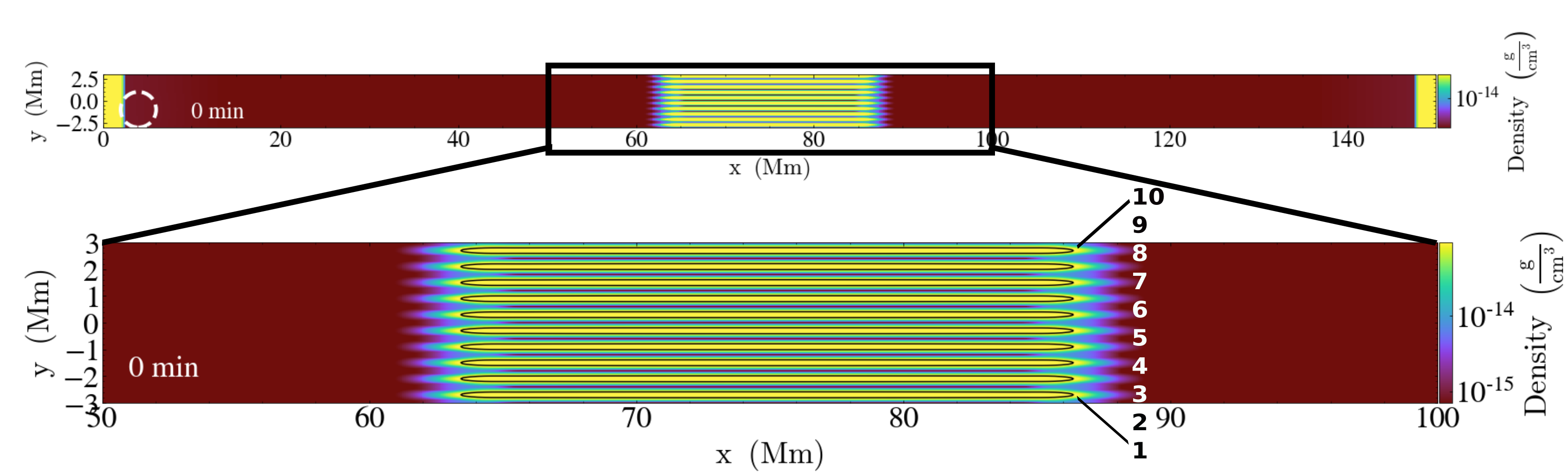}
   \caption{Density plot of the domain (at $t=0$), with the dashed circle denoting the area where we increase the energy (source region). We note that the variation along $x$ is actually  a variation along magnetic field lines, which have the prescribed shape as shown in Fig.~\ref{fig:mag_topology}. The bottom of the figure represents a cut-out focusing on the area of the threads located at the center of the domain. The contour lines mark the area with number density higher than 5$\times$10$^{10}$\,cm$^{-3}$ (where the plotted density is $\rho=1.4m_pn_H$). We label the threads with the numbers \textit{1}--\textit{10} and refer to them as such throughout the paper.}
    \label{fig:domain}
\end{figure*}

\begin{figure}
\centering
\includegraphics[width=0.48\textwidth]{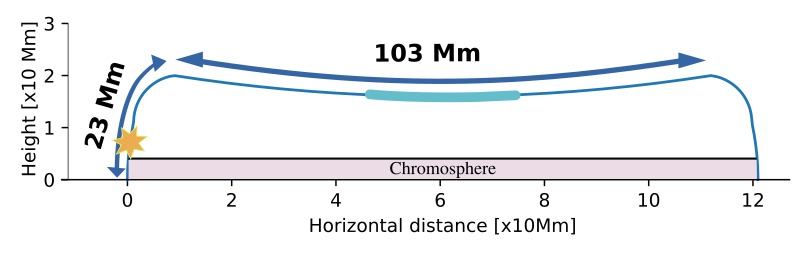}
  \caption{Geometry of the magnetic arcade, with a prominence positioned at the center of the arcade and a star marking the position of the additional source of energy. The pink area at the bottom indicates the chromosphere (and photosphere) where the coronal arcade is rooted. This prescribed field line shape is repeated along the entire $y$ direction of the arcade shown in Fig.~\ref{fig:domain}.}
\label{fig:mag_topology}
\end{figure}

   For this study, we work in the 2D domain, as presented in Fig.~\ref{fig:domain}. The magnetic arcade contains a dip at the centre and is symmetric about the midpoint (Fig.~\ref{fig:mag_topology}). The setup we envision uses a similar geometry as in previous 1D hydrodynamic models \citep{Mok1990, Antiochos1999, Karpen2001, Xia2011, Zhang2012, Luna2012c, Zhang2013, Zhou2014, Zhang2020}, as presented in Fig.~\ref{fig:mag_topology}. In contrast to all previous 1D works on prominence formation and oscillations, we repeat Fig.~\ref{fig:mag_topology} in the plane shown in Fig.~\ref{fig:domain}. That makes our simulation a full 2D MHD simulation in which we artificially insert a threadlike appearance, as described in what follows. The legs and shoulders of the prominence have the shape of a circular arc (each of a different radius, approximately 50 and 9\,Mm, respectively). The central part is represented with a concave upwards parabola that connects the two opposite sides and represents the central dipped arcade part, where the prominence settles. The height of the arcade is 20\,Mm, leg and shoulder (on each side) are 23\,Mm long and the central part is 103\,Mm in length. In total, the domain is approximately 150\,Mm long and 6\,Mm wide (Fig.~\ref{fig:domain}). To create a gravitationally stratified atmosphere, we used the following function to describe the temperature change with the height, $z,$ (where the height corresponds to the $x$ in our domain):
   
\begin{equation}
    T(z) =  T_{pho} + \frac{1}{2} (T_{cor}-T_{pho})\Bigg[1 + tanh \bigg( \frac{z-h_{tra}}{w_{tra}} \bigg) \Bigg] \,,
    \label{eq:temperature}
\end{equation}
   
   \noindent Equation~(\ref{eq:temperature}) is similar as in \cite{Xia2011} and in \cite{Zhou2017}. Here, $T_{pho}$ is the temperature of the photosphere, 6000\,$\mathrm{K}$ and $T_{cor}$ is the coronal temperature of 1\,MK, while $h_{tra}$ and $w_{tra}$ are the height and width of the transition region (TR). To get the values of pressure and density, we use hydrostatic equilibrium \citep[for details see][]{Xia2011}. Since we do not study the formation process, the thread-like structure is artificially created by increasing the density at the centre, with regards to the background density. We increase the density by keeping the pressure constant \citep{Terradas2015, Zhou2017}. There are in total ten threads whose area is initially 25\,Mm long and 200\,km wide. The number of threads and their dimension are chosen such that the filling factor (volume ratio of a certain material to the total structure volume) is in the range of observational values \citep[cf.][]{Labrosse2010}. To reach a magnetohydrostatic (near-) equilibrium, before introducing perturbations, we let the system relax for about an hour. After that time the maximal velocity of the system is less than a km\,s$^{-1}$. The numerical relaxation has slightly reduced the threads' original length and increased their width. If we define the threads as interior to the contour lines of number density, $n_H>5\times$10$^{10}$\,cm$^{-3}$ (as shown in Fig.~\ref{fig:domain}, where $\rho=1.4m_pn_H$, and $m_p$ is the proton mass), they reach a length of about 23\,Mm and a width of 230\,km (Fig.~\ref{fig:domain}). At the end of the relaxation phase temperature, $T$ and number density, $n_H$ in the corona reach values of 4\,$\mathrm{MK}$ and 2.7$\times10^8$\,cm$^{-3}$. At the endpoints of our domain (representing bottom of chromosphere), $T$ and $n_H$ are about 7000\,$\mathrm{K}$ and 7.8$\times10^8$\,cm$^{-3}$. 

\subsection{Numerical strategy} \label{subsec:numerics}

   The shock wave that we simulate corresponds to coronal bright fronts frequently seen in observations. Most often they are a result of a CME or a flare and have velocities of 200-500\,km\,s$^{-1}$, but are also able to reach values of $\approx$1400\,km\,s$^{-1}$ \citep[][and references therein]{Long2017}. The mechanisms of formation of coronal shock waves are still not precisely known. The most accepted theory is that the trigger mechanism is magnetic reconnection. Reconnection releases large energies that are transported toward the chromosphere along magnetic field lines \citep{Shibata&Magara2011}. As a result, there is an increase in gas pressure that sets the plasma moving toward the corona, against the act of the gravitational force. Since we are limited with a simple adiabatic setup and an ideal MHD we decided to simulate a source region simply as a confined area of increased energy (i.e. pressure). As the goal was not to study the formation of a shock wave in detail, this approach allows a more realistic but relatively simple view of the source of the oscillations. We define the spatial and temporal scales of such energy release based on the commonly used Gaussian profile. Due to the small timescale  in which we increase the energy, a shock wave results. A similar approach of using a pressure pulse can be also found in the work of \cite{Pascoe2009}. Numerically, we introduce an additional source to the energy equation. The source is described with the following equation:
   
\begin{equation}
    S(x,y,t)=S_0 \exp\Bigg[-\frac{r^2}{R^2}-\frac{(t-t_{peak})^2}{t_{scale}^2}\Bigg] \,,
    \label{eq:source}
\end{equation}

   \noindent where $r=[(x-x_s)^2+(y-y_s)^2]^{1/2}$ is the radial distance from the source centre ($x_s=4$ Mm and $y_s= -1$ Mm); $t_{peak}$ is the time when the source is maximised, which is 85\,s after the end of the relaxation phase. Moreover, the end of the relaxation phase is the moment we take to be $t=0$. $t_{scale}$ is the duration of the source (25.5\,s). In addition, $S_0$ determines the amplitude of the source, calculated as the energy of the source per unit volume (taking into account that the $x$ coordinate actually corresponds to the vertical coordinate of the source) and time. We analyzed two cases with different amplitudes, where we changed the energy of the source (10$^{26}$\,erg and 2$\times$10$^{26}$\,erg, from now on referred to as low\_ampl and high\_ampl cases, respectively) and kept the same volume and time during which we introduce the source (50.27\,Mm$^3$ and 25.5\,s). According to the results of a statistical analysis of microflares \citep{Hannah2008}, in both cases, we are dealing with relatively weak flares. We add this additional energy to the system when any existing velocities in the domain drop below 1\,km\,s$^{-1}$ (after the relaxation phase, at $t=0$). With this type of trigger, we induce oscillations without preselecting a specific polarisation.

   Conditions in the corona imply a region of low thermal pressure and relatively high magnetic pressure (i.e. for plasma beta is valid $\beta$ $\ll$ 1, where $\beta=2p/B^2\mu_0$). We do not include any effects of resistivity or viscosity, therefore, the magnetic field lines guide the plasma, and the following ideal MHD equations describe its dynamics:

\begin{equation}
\label{eq:MHD1}
    \pdv{\rho}{t}+\nabla \cdot (\rho \textbf{v}) = 0 \,,
\end{equation}

\begin{equation}
\label{eq:MHD2}
    \pdv{\rho \textbf{v}}{t} + \nabla \cdot \Bigg(\rho \textbf{vv} + p_{tot}\textbf{I}-\frac{\textbf{BB}}{\mu_0} \Bigg) = \rho \textbf{g} \,,
\end{equation}

\begin{equation}
\label{eq:MHD3}
    \pdv{e}{t} + \nabla \cdot \Bigg(e\textbf{v}-\frac{\textbf{BB}}{\mu_0}\cdot \textbf{v} + \textbf{v}p_{tot} \Bigg)=\rho \textbf{g}\cdot \textbf{v} + S \,,
\end{equation}
    
\begin{equation}
\label{eq:MHD4}
    \pdv{\textbf{B}}{t} + \nabla \cdot (\textbf{vB} - \textbf{Bv})=0 \,,
\end{equation}

   \noindent where $p_{tot}=p+\frac{B^2}{2\mu_0}$ is the total pressure, $\textbf{g}$ is gravity whose $x$-component corresponds to a fixed vertical gravity component that is locally projected along the prescribed field line shape, where the value of gravity at the solar surface is 274\,m\,s$^{-2}$. The geometry of the magnetic arcade determines the distribution of the field-aligned gravity component.   Also, $\rho$, $\textbf{v}$ $e$, and $\textbf{B}$ are plasma density, velocity, energy density and the magnetic field, respectively; $S$ is the source term from Eq.~(\ref{eq:source}) that is adopted as in  Eq.~(\ref{eq:MHD3}). Since we do not focus on the formation phase of the prominence threads nor the damping effects of the induced prominence oscillations, the setup is adiabatic. This allows us to draw clear inferences on cause and effect, and to analyse the temporal dynamic behaviour of the multi-threaded, shock-impacted prominence.

   \noindent In order to solve Eqs.~(\ref{eq:MHD1})-(\ref{eq:MHD4}) we used an open-source MHD simulation code, the  MPI\ Adaptive Mesh Refinement Versatile Advection Code (MPI-AMRVAC\footnote{http://amrvac.org/}) \citep{Keppens2012, Porth2014, Chun2018, Keppens2021}. We carried out the time discretisation with a five-step (strong stability preserving) fourth-order Runge-Kutta method \citep{Spiteri2002}, while for the spatial discretisation, we employed a Harten-Lax-van Leer (HLL) approximate Riemann solver \citep{Harten1983} combined with a second order shock-capturing slope limiter \citep{vanLeer1977, Woodward1984}. To ensure stability, the courant number we used is 0.8. Furthermore, to resolve the fine structure of the prominence, we took advantage of the adaptive mesh refinement capabilities of AMRVAC. We used four levels of AMR (including the base grid level)  to obtain a dynamic grid that effectively consists of 4160$\times$800 cells, which allows us to resolve lengths of 36\,km\,$\times$\,7.5\,km. The refinement criteria are based on density, hence, the finest mesh refinement happens around the threads (and the boundary region). For the boundary conditions along the $x$ coordinate (footpoints), we adopted line-tied boundary conditions. Pressure and density are fixed to values corresponding to the values in the photosphere (according to the hydrostatic equilibrium). The TR is included on both footpoint regions of the simulated arcade. Although the temperature can shift rapidly around that region, the analysis here is focused on the interaction of the shock wave with the threaded prominence. For the velocity, we used the reflection condition and the magnetic field is extrapolated, which means we used the values in the computational domain to calculate values in the ghost cells. For simplicity, we used periodic boundary conditions along the $y$-coordinate. The magnetic field is uniform initially in the $x$ direction with the initial value of 10\,G. The corresponding plasma $\beta$ changes from about 40 in the chromosphere to 0.009 in the corona. The 2D nature of the prescribed magnetic field is such that it can bend in the direction across the arcade, but it cannot alter the prescribed shape given in Fig.~\ref{fig:mag_topology}. To further improve the efficiency and accuracy, the code has the ability of a magnetic field splitting strategy according to \cite{Tanaka1994}. With it, the magnetic field is split into a time-invariant part which is handled exactly (background magnetic field) and the remaining part is the perturbed magnetic field for which we solve the equations. To control the divergence of the magnetic field, the code has more than one approach. In our case, a combined approach is used, where different source terms are added to the governing equations in order to diffuse \citep{Keppens2003} or control numerical errors \citep{Powell1999}.

   \section{Results} \label{sec:results}

   The prominence material is defined as plasma with a number density higher than 5$\times$10$^{10}$\,cm$^{-3}$. With that threshold, we locate the moving and thermodynamically adjusting ten prominence threads during the entire simulation. At $t=0$ this is shown in Fig.~\ref{fig:domain}. After the relaxation phase, the average density of the threads is about 240 times greater than the plasma density of the coronal domain (thus, we focus specifically on the core region of the threads). After adding energy into the system (as described in Sect.~\ref{sec:method}), a resulting shock wave perturbs the threads. In this section we start by describing the properties of the low\_ampl case (reference case) and then we make a comparison with the high\_ampl case. The time it takes for the disturbance from the source to reach the threads is less than 3\,min. That corresponds to a velocity of the disturbance of approximately 400\,km\,s$^{-1}$, which matches with often observed values of EUV waves \citep{Gallagher2011,Long2017}. The initially circular wave hits the threads that are then set into motion. However, the wave itself is also partially reflected off the lower-lying TR, so the wave is more complex than a simple circular wavefront. We follow the thread oscillations for $\sim$140 min, during which a few cycles of longitudinal oscillations are measurable. For further reference, the approximate value of the sound velocity in the corona is 327\,km/s and of the Alfv{\'{e}}n velocity is 1066\,km/s. 

\begin{figure}
    \centering
    \includegraphics[width=0.45\textwidth]{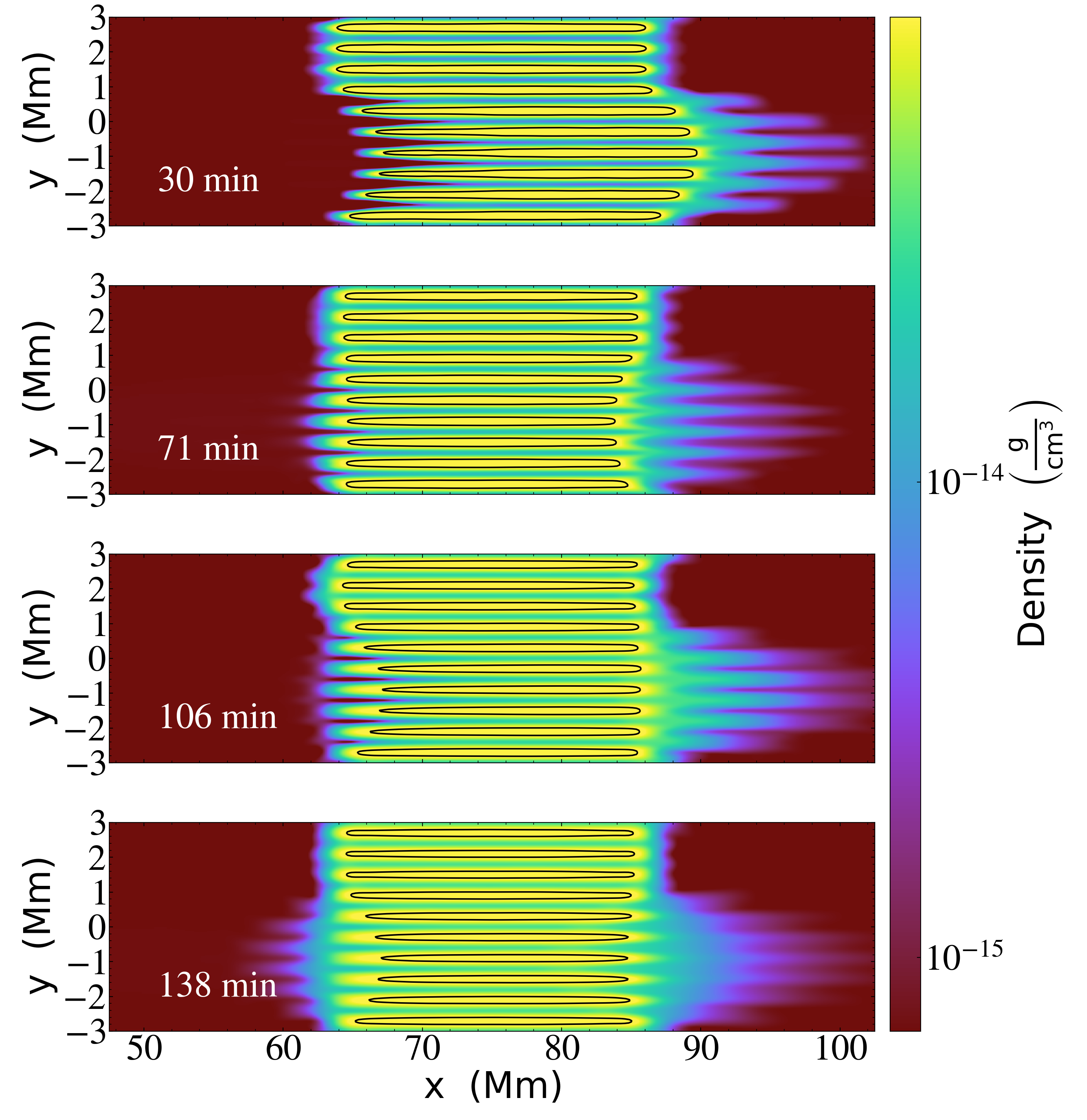}
    \caption{Snapshots of oscillations with contour lines marking the area with number density higher than 5$\times$10$^{10}$\,cm$^{-3}$. We note that we show the central part of the arcade only, similar as in the bottom part of Fig.~\ref{fig:domain}}
    \label{fig:snapshots}
\end{figure}

\subsection{Longitudinal oscillations}

   As the shock wave hits the threads, it encounters a density gradient (the so-called prominence corona transition region, PCTR), and reflects off of it. The time it takes for the initial shock wave to pass the (coronal) area between the threads and to reach the right TR is about 6\,min. There, it again encounters a sharp density and temperature gradient (namely the rightmost TR) and it reflects back. Based on the research on EUV waves \citep[][and references therein]{Long2017}, we know that they exhibit properties of reflection, refraction, and transmission. This means that in our simulation, the original wave energy is also partially transmitted down into the chromospheric regions at both sides. Despite the source region we impose and the multiple reverberations happening in the domain, the threads still maintain a relatively coherent motion. 

Figure~\ref{fig:snapshots} shows four different moments during their induced oscillation (with contours marking the threads). After about 30\,min after introducing the source, the threads reach their maximum deviation from their initial equilibrium position. As expected, the source region moves the threads mainly along the magnetic field lines. All the field lines have the same curvature as prescribed by Fig.~\ref{fig:mag_topology}, but the threads on different field lines experience a slightly different impact (in timing and the angle of attack) from the shock wave. Thread \textit{4} was hit directly; consequently, it has the most significant deviation from its initial position and it reaches the highest temperature as a result of the compression it experiences. At the same time, threads \textit{8, 9,} and \textit{10} were not on the direct path of the shock wave and experienced less intense changes (first snapshot of Fig.~\ref{fig:snapshots}). At 71\,min, when the threads reach their maximum displacement, we notice on the left that it is not strikingly pronounced, while the next maximum displacement at 106\,min on the right side is more discernible. The last snapshot at $t=138$\,min corresponds to the moment when the threads are almost settled at their new equilibrium position. As seen in Fig.~\ref{fig:snapshots}, the impacting wave redistributes mass (and changes the temperature) of the threads markedly, but we wish to identify the instantaneous thread position based on a quantitative measure: the contour corresponding to the original 5$\times$10$^{10}$\,cm$^{-3}$. It can then be seen in Fig.~\ref{fig:snapshots} that the density (and temperature) changes are noticeable on a larger area than the one outlined by the thread.

\begin{figure}
    \centering
    \includegraphics[width=0.45\textwidth]{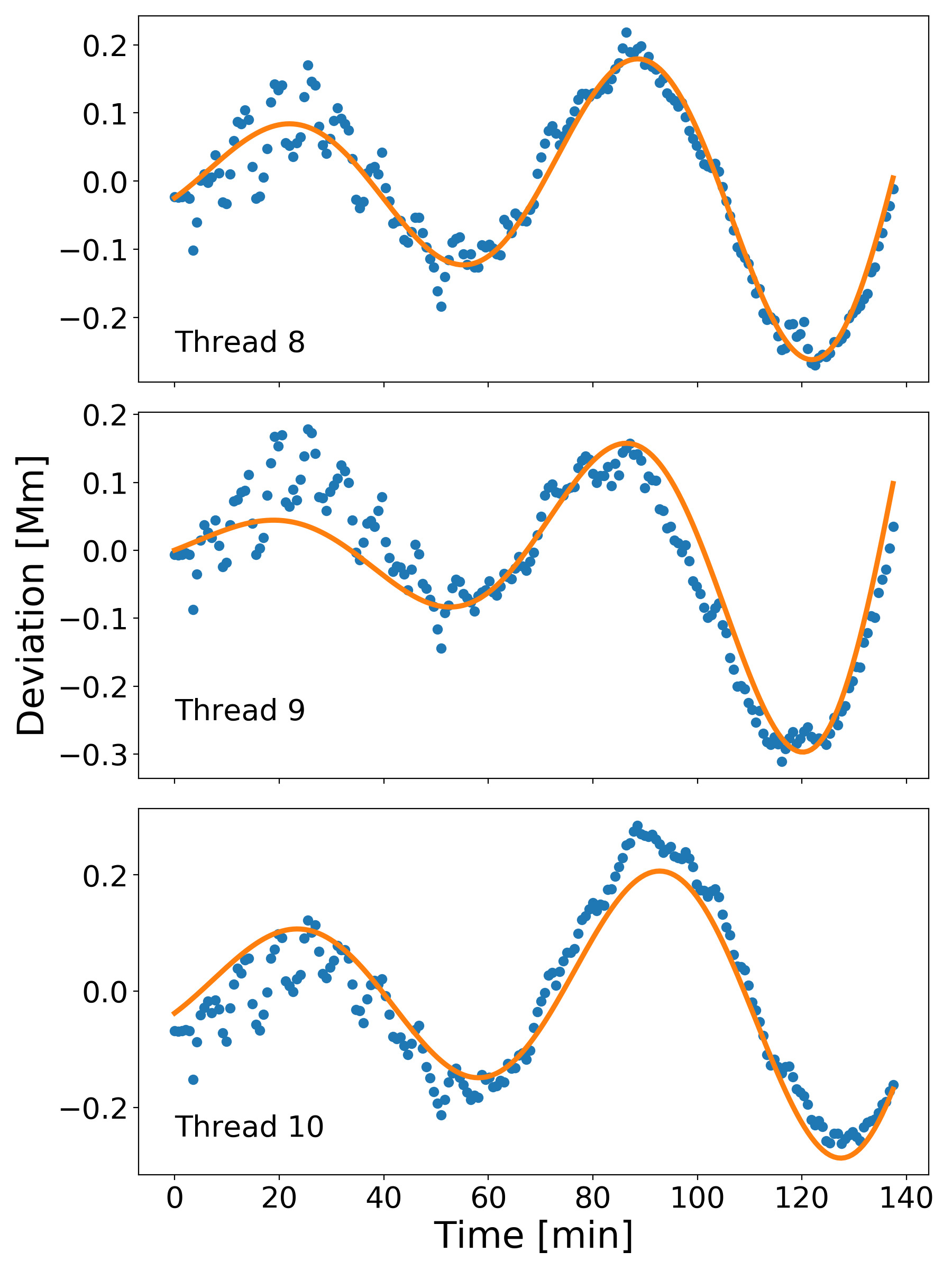}
    \caption{Evolution of motion of the CM along the $x$ direction, $d_x^{CM}(t)$ of threads \textit{8}--\textit{10} with the orange line representing the fit function described with Eq.~(\ref{eq:fitting_func}).}
    \label{fig:CMx8_10}
\end{figure}

   To follow the distinct motion of each thread, we calculated the coordinates of the centre of mass (CM), its $x$ and $y$ components in time: 

\begin{equation}
     d_x^{CM}(t) = \frac{\int_{\mathcal{A}} x(t)\rho(t,\Vec{x}) dA}{\int_{\mathcal{A}} \rho(t,\Vec{x}) dA} \,,
\end{equation}
\begin{equation}
     d_y^{CM}(t) = \frac{\int_{\mathcal{A}} y(t)\rho(t,\Vec{x}) dA}{\int_{\mathcal{A}} \rho(t,\Vec{x}) dA} \,,
\end{equation}
 
   \noindent where $\mathcal{A}$ denotes the area of each thread outlined by the number density threshold value of 5$\times$10$^{10}$\,cm$^{-3}$ and $\textbf{x}$ is the position vector. From previous works \citep{Hyder1966,Vrsnak2007,Luna2012b}, we know that in a dipped magnetic arcade, we can expect the prominence to behave as a harmonic oscillator. The measured displacement from the initial position in the $x$ direction, $d_x^{CM}$ can be fitted with a decaying sinusoidal function $f(t)$:

\begin{equation}
    f(t)=A_0e^{-t/\tau}sin(\omega t + \phi) + b \,,
    \label{eq:fitting_func}
\end{equation}

   \noindent where $A_0$ is the amplitude, $\tau$ is damping time, $\omega$ is the frequency, $\phi$ is the phase, and $b$ is the equilibrium position (to fit the average velocity of each thread we used the same type of sinusoidal function). The results are given in Table~\ref{table:displ} and \ref{table:vel}. The displacement periods of each thread are in the range of 66 to 85\,min (Table~\ref{table:displ}), with the velocity periods in the range of 62 to 73\,min (Table~\ref{table:vel}). Since the ten threads represent separate entities that are changing their total masses (i.e. transferring some to coronal material as judged from the density contour alone), they show discrepancies with what we would expect if the prominence was modelled as a plain solid body. The first seven threads \textit{(1}--\textit{7)} exhibit an expected behaviour: they are pushed and, as a result, they oscillate and eventually settle back to their equilibrium position. They show damped oscillations (Table~\ref{table:displ}) with the ratio of damping time and period ($\tau$/$P$) indicating strong damping, where thread \textit{7} is perturbed to a lesser extent. However, threads \textit{8, 9,} and \textit{10} exhibit amplification in their displacement and subsequently in their velocity of oscillation (negative damping times). We show this in Fig.~\ref{fig:CMx8_10}. To adequately describe each thread, we calculated average values of:
\begin{equation}
    v_i(t)= \frac{\int_{\mathcal{A}} v_i(t,\Vec{x})\rho(t,\Vec{x}) dA}{\int_{\mathcal{A}} \rho(t,\Vec{x}) dA} \,,
    \label{eq:average_velocity}
\end{equation}  
\begin{equation}
    B_i(t)= \frac{\int_{\mathcal{A}} B_i(t,\Vec{x}) dA}{\int_{\mathcal{A}} dA} \,,
\end{equation}    
\begin{equation}    
    p(t)= \frac{\int_{\mathcal{A}} p(t,\Vec{x}) dA}{\int_{\mathcal{A}} dA} \,,
\end{equation}    
\begin{equation}   
    T(t)= \frac{\int_{\mathcal{A}} T(t,\Vec{x}) dA}{\int_{\mathcal{A}}dA} \,,
\end{equation}
   
\begin{figure}
    \centering
    \includegraphics[width=0.45\textwidth]{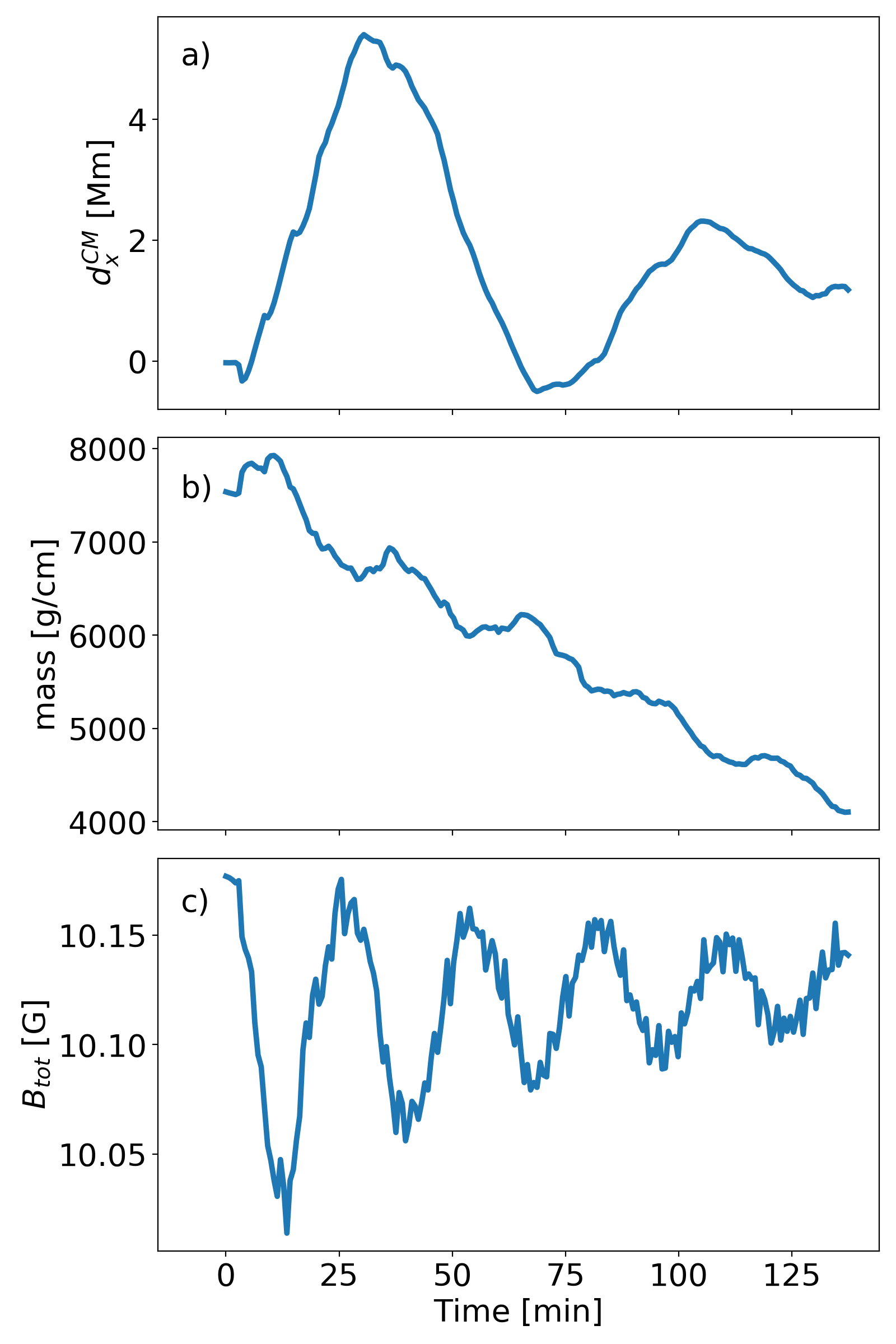}
    \caption{Evolution of motion of the CM along the $x$ direction, $d_x^{CM}(t)$ and the average values of the total magnetic field and mass during oscillations of thread \textit{4}.}
    \label{fig:CMx_Btot_mass_thr4}
\end{figure}

\begin{figure}
    \centering
    \includegraphics[width=0.45\textwidth]{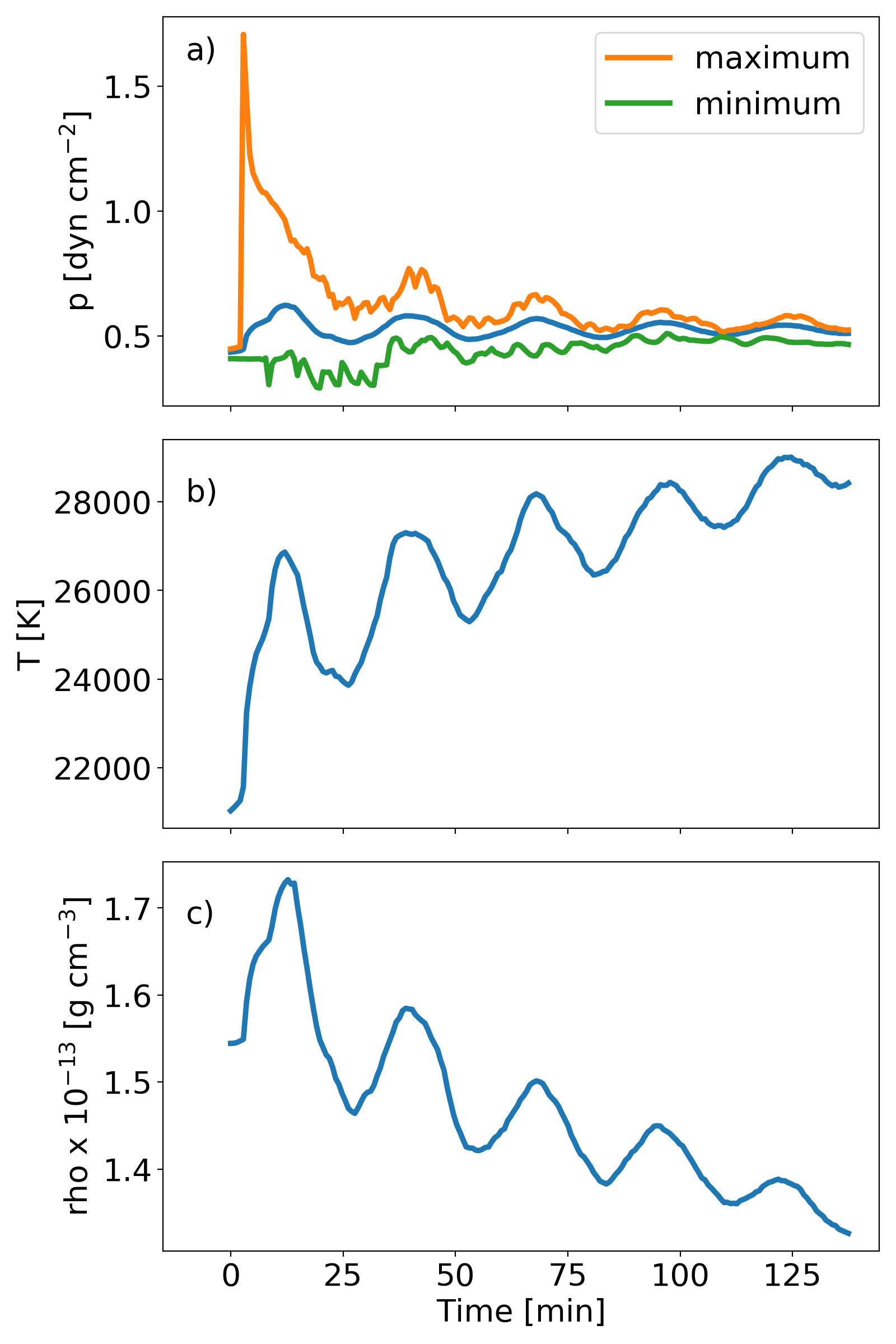}
    \caption{Evolution of the gas pressure, temperature and density during oscillations of thread \textit{4}.}
    \label{fig:p_t_rho_thr4}
\end{figure}

   \noindent where index $i$ stands for the $x$ and $y$ components of the velocity and the magnetic field. The timelines of these average values are presented in Fig.~\ref{fig:CMx_Btot_mass_thr4} and~\ref{fig:p_t_rho_thr4} (where the density is the averaged mass over the volume, and the average total magnetic field, $B_{tot}$, is the square root of the sum of the squared components, $B_x$ and $B_y$). Besides the value averaged over area, plot a) of Fig.~\ref{fig:p_t_rho_thr4} also shows how the minimum (green line) and maximum (orange line) values found inside the area of thread \textit{4} change. We can notice that those values are, particularly in the beginning, strongly influenced by the impact of the shock wave. Later on, they follow the main smooth evolution shown by the area-averaged value, however, they are more irregular. For that reason, we opted to use the area-averaged values as parameters describing the evolution of the threads oscillation. We took thread \textit{4} as a representative, since it has the highest deviations and considering other threads show a similar behaviour. Each thread, before the oscillations start, has a mass of 7.54$\times$10$^{3}$\,g/cm. Summing all the threads, the complete prominence has a mass of 7.54$\times$10$^{4}$\,g/cm. From panel b) in Fig.~\ref{fig:CMx_Btot_mass_thr4} we can see that the mass of each extracted thread continuously decreases during oscillation, an effect that is visually represented in Fig.~\ref{fig:snapshots}, as some matter is now counted as coronal mater instead. At the end of the oscillation, the masses of each thread are in the range of (4.09--4.87)$\times$10$^{3}$\,g/cm (corresponding to masses of thread \textit{4} and \textit{7}, respectively). The magnetic field after the relaxation phase has a value of 10.18\,G. We can see from panel c) of Fig.~\ref{fig:CMx_Btot_mass_thr4} that the large scale periodic motion of the magnetic field follows the changes in pressure and density (in antiphase), shown in Fig.~\ref{fig:p_t_rho_thr4}. When there is  compression and an increase in the average density of the thread, the magnetic field of that thread experiences a drop in value. When there is a decrease in the average thread density, the magnetic field starts increasing again and returning to its initial value. Thread \textit{4} experiences the greatest changes in the magnetic field (up to 0.16\,G in the $x$ component), while threads \textit{8}, \textit{9,} and \textit{10} are the least perturbed. It's also important to note that the changes seen in the total magnetic field are primarily due to the longitudinal $B_x$ changes. \\ We conducted a Fourier analysis of different parameters to extract significant periods of oscillation in the $x$ and the $y$ components. We sampled the data every 42.5\,s, which makes certain periods (below 0.7\,min) indistinguishable. The $x$ component of the magnetic field has a period of low-frequency motion of 27.4\,min (same as pressure, density, and temperature) with high-frequency oscillations predominantly in the range of 1.5 to 16\,min. In panel b) of Fig.~\ref{fig:p_t_rho_thr4} we can also notice that the temperature of thread \textit{4} shows the same low-frequency oscillations as the magnetic field, but with a steady increase in value. This increase in the average temperature is evident in all the threads and is just representing the work/heating done by compression. Each thread starts oscillating with a typical initial temperature of 21035\,$\mathrm{K}$. At the end of oscillation, the temperature of each thread is in the range of 25679 (thread \textit{9}) to 28450\,$\mathrm{K}$ (thread \textit{4}). Thread \textit{4} reaches the highest temperature, while threads \textit{8, 9,} and \textit{10} have the lowest increase. 

   For the high\_ampl case, all the parameters experience stronger perturbations. The initial displacement doubles. The initial velocity amplitude for threads \textit{1}--\textit{6} also doubles, while it doesn't change significantly for threads \textit{7}--\textit{10}. Damping time is shorter for the high\_ampl case, again more obvious for threads \textit{1}--\textit{6} than for threads \textit{7}--\textit{10}. At the same time, the period of oscillation becomes longer for threads \textit{1}--\textit{6} in the high\_ampl case than in the low\_ampl case, while the velocity periods do not change notably. The threads transfer more mass to the corona than in the low\_ampl case. At the end of oscillation in the high\_ampl case, the threads have their mass in the range (3.55--4.94)$\times$10$^{3}$\,g/cm (corresponding to masses of thread \textit{4} and \textit{7}, respectively). While the temperature range in which the threads of the high\_ampl case end up varies from 25803\,$\mathrm{K}$ up to 32253\,$\mathrm{K}$ (corresponding to average temperatures of thread \textit{9} and \textit{4}, respectively). The low-frequency oscillations seen in magnetic field, pressure, density, and temperature in Fig.~\ref{fig:CMx_Btot_mass_thr4} and ~\ref{fig:p_t_rho_thr4} also appear for the high\_ampl case, where the measured period is 28.2\,min. The high-frequency oscillations seen in the $x$ component of the magnetic field appear in approximately the same range as in the low\_ampl case (1.5 to 16\,min).

\begin{table}
\caption{Values of amplitude, period and damping time of the $x$ component of the CM coordinate, $d_x^{CM}(t)$ for each thread in the low\_ampl case.}  
\label{table:displ}      
\centering                          
\begin{tabular}{c c c c}        
\hline\hline                 
Thread & Amplitude & $\tau$ & Period \\    
  &  [Mm] & [min] & [min] \\  
\hline\hline                        
  1 & 1.14 & 111.46 & 77.08  \\      
  2 & 2.80 & 90.93 & 81.25 \\
  3 & 4.24 & 76.50 & 83.63 \\
  4 & 4.72 & 73.49 & 84.76 \\
  5 & 3.75 & 84.88 & 81.88 \\ 
  6 & 2.57 & 82.61 & 83.74 \\
  7 & 0.45 & 521.30 & 70.13 \\
  8 & 0.07 & -88.02 & 66.61 \\
  9 & 0.03 & -53.19 & 67.48 \\
  10 & 0.09 & -105.45 & 69.31 \\
\hline
\hline
\end{tabular}
\end{table}

\begin{table}
\caption{Values of amplitude, period and damping time calculated by fitting the longitudinal velocity oscillations for each thread in the low\_ampl case.}
\label{table:vel}
\centering 
\begin{tabular}{c c c c}
\hline\hline
Thread & Velocity ampl. & Velocity $\tau$ & Velocity period \\ 
  &  [km s$^{-1}$] & [min] & [min] \\  
\hline\hline
1 & 1.42 & 96.22 & 69.45 \\
2 & 3.34 & 81.88 & 71.77 \\
3 & 4.87 & 74.39 & 71.82 \\
4 & 5.30 & 72.54 & 71.96 \\
5 & 4.42 & 79.35 & 71.88 \\
6 & 2.96 & 77.03 & 72.28 \\
7 & 0.58 & 437.90 & 67.84 \\
8 & 0.10 & -114.51 & 65.72 \\
9 & 0.08 & -92.80 & 62.97 \\
10 & 0.11 & -105.53 & 70.71 \\
\hline 
\hline
\end{tabular}
\end{table}

\subsection{Transverse oscillations}

\begin{figure}
    \centering
    \includegraphics[width=0.48\textwidth]{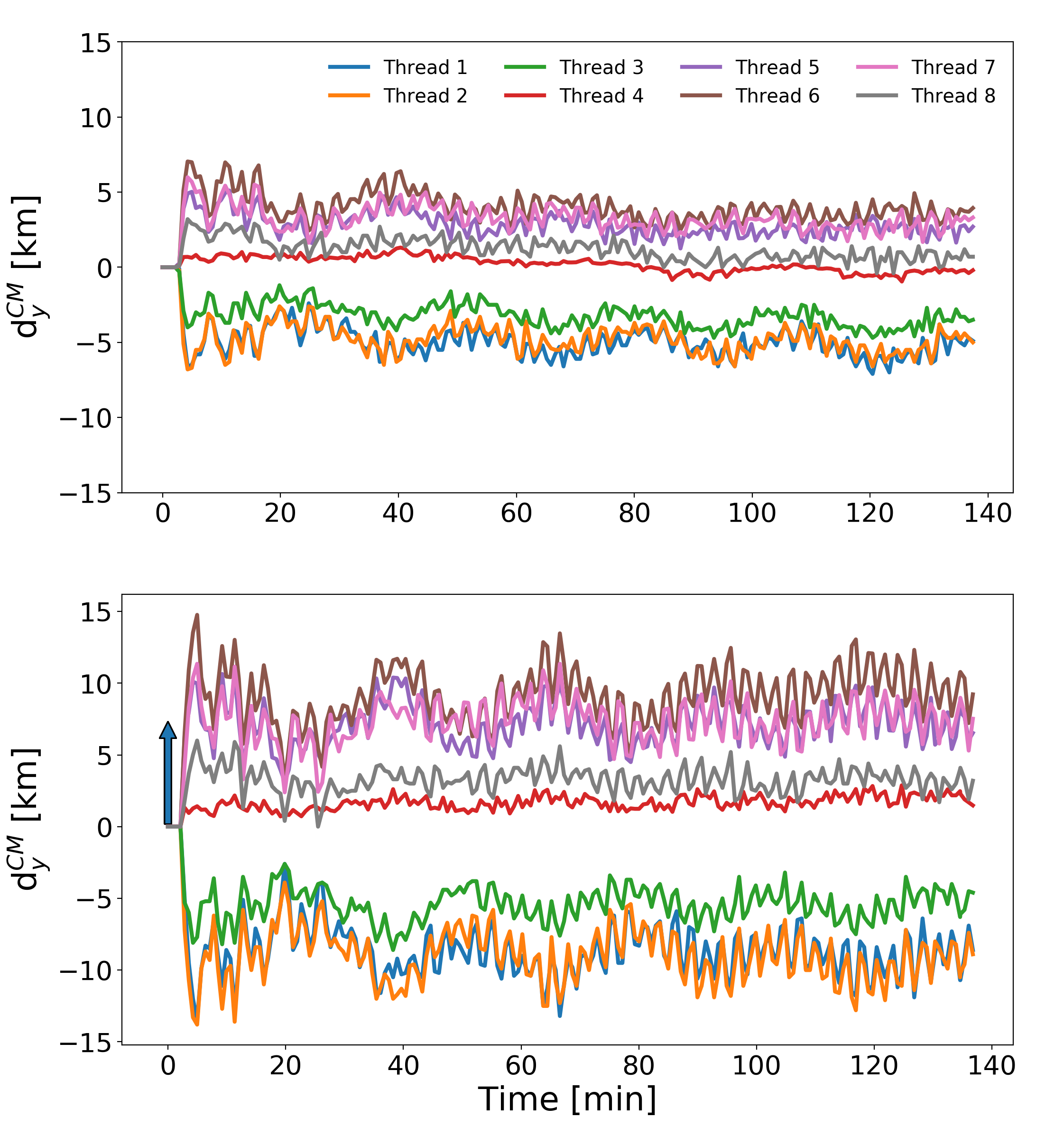}
    \caption{Evolution of motion of the CM coordinate in the $y$ direction of threads \textit{1}--\textit{8} (transverse displacement) in respect to their initial position. The top plot shows d$_y^{CM}$ for low\_ampl case and the bottom plot for the high\_ampl case. The arrow on the bottom plot marks the size of the cell in the $y$ direction (7.5\,km).}
    \label{fig:dy}
\end{figure}

   Longitudinal oscillations markedly dominate the motion of the threads with no obvious displacements in the direction that is perpendicular to it. However, upon further examination, oscillations in the transverse direction are also noticeable. The resulting motions are very small, and the actual displacement is the same order of our grid size in the $y$ direction for low\_ampl case. Also, the velocities are less than a km/s and not easily perceptible when viewing it on the large-scale domain. Nonetheless, this type of motion becomes increasingly more distinct as we increase the amplitude of the source. The transverse oscillations observed in this simulation show a certain regularity and are important in the interpretation of the total motion of each thread. In Fig.~\ref{fig:dy} we show the first eight threads and their $y$ component of the CM, $d_y^{CM}$. We decided to leave out the remaining three threads as they show an even smaller deviation. Thread \textit{4} is displaced slightly upwards, which is understandable considering it has its $y$ coordinate 100\,km shifted from the $y$ coordinate of the centre of the source. Every thread, with a transverse $y$ coordinate displaced with respect to the centre of the source region (up to and including thread \textit{8}) is pushed slightly sideways (to the left). Threads located to the right of the centre of the source region are pushed rightwards, including top threads \textit{9} and \textit{10}. The period of the low-frequency oscillations of $d_y^{CM}$ is 27.4\,min for threads \textit{1} to \textit{4} and threads \textit{9} and \textit{10}. Threads \textit{5} to \textit{8} have a slightly longer period of 32.9\,min. This trend gets more obvious as the amplitude of the source is increased. The $v_y$ component has only the high-frequency oscillations (the low-frequency motion is not evident). On top of the low-frequency periodic motion, quite noticeable in Fig.~\ref{fig:dy} are also high-frequency oscillations. The high-frequency oscillations for the studied parameters, $d_y$, $v_y$, $B_x$, and $B_y$ generally appear in the range of about 1.5 to 16\,min. 
   
   For the high\_ampl case, we again notice the same behaviour, with greater transverse deviation from their equilibrium position (bottom panel of Fig.~\ref{fig:dy}). In both the low\_ampl and high\_ampl cases, the strength of our source region is not relatively high. However, it is already noticeable that by increasing its strength, the oscillation in the $y$ direction, while small, is still undoubtedly increasing. The periodicities are mostly of the same values in both cases, $d_y^{CM}$ has a low-frequency oscillation of period 28.2\,min for all threads. Again, $v_y$  has no low-frequency oscillations and the high-frequency oscillations for $d_y$, $v_y$, $B_x$, and $B_y$ appear in the same period range as in the low\_ampl case (1.5 to 16\,min).

\section{Discussion} \label{sec:discussion}

   Most of the previous  works on the topic of prominence oscillations were carried out on the basis of oscillations that were either parallel to the background magnetic field (longitudinal oscillations) or perpendicular to it in the vertical direction (vertical transverse oscillations). These oscillations were simply induced by introducing velocity perturbations, which is directly rigidly displacing the prominence as a whole. Here, we focus on oscillations in the curved surface that is formed by the magnetic field of a realistic arcade. This allows us to handle the coexistence of both longitudinal and transverse oscillations, hereby, also studied in an actual multi-threaded prominence topology. Furthermore, the oscillations are induced with a realistic source region. The high resolution we employ makes it possible to study the dynamic response to the perturbation of the fine prominence structure. Due to the ideal MHD assumption and the sideways periodic boundary conditions, we have simplified the evolution to adiabatic, but otherwise easily controlled, numerical experiments. Even so, the results presented here offer valuable insights into the dynamics of fine-structured solar prominences and we go on to discuss the physical aspects that will also have ramifications for future simulations in 3D and even more realistic ones.

\subsection{Longitudinal oscillations and the restoring force}

   Both cases of different source amplitude analyzed here show the same type of behaviour, in the sense that thread \textit{4} has the longest period of the associated longitudinal oscillation and the threads around it show consequently smaller period. Such behaviour is a result of the position of each thread as compared to the position of the centre of the source region. The forces acting along the field line direction are the pressure gradient force and the projected gravity component. Because the direction of motion is dominantly parallel to the magnetic field, the Lorentz force in the $x$ direction is essentially vanishing. At the initial moment, after the relaxation phase and before the shock wave hits the threads, a small compression force in the $x$ direction exists ($<$\,1$\times$10$^{-10}$\,dyne/cm$^{3}$), however, it is considerably smaller in respect to the forces caused by the shock wave. Thus, we achieved satisfyingly small initial values of velocities in the domain and an approximate equilibrium state. By introducing the source in a self-consistent way, the resulting shock wave shifts the system of threads from their initial equilibrium position slightly along the field lines they reside in, for about 1.2\,km for thread \textit{8} to 1.5\,Mm for thread \textit{4} (which explains the need for $\phi$ and $b$ parameters in Eq.~(\ref{eq:fitting_func})). 

\begin{figure}
    \centering
    \includegraphics[width=0.45\textwidth]{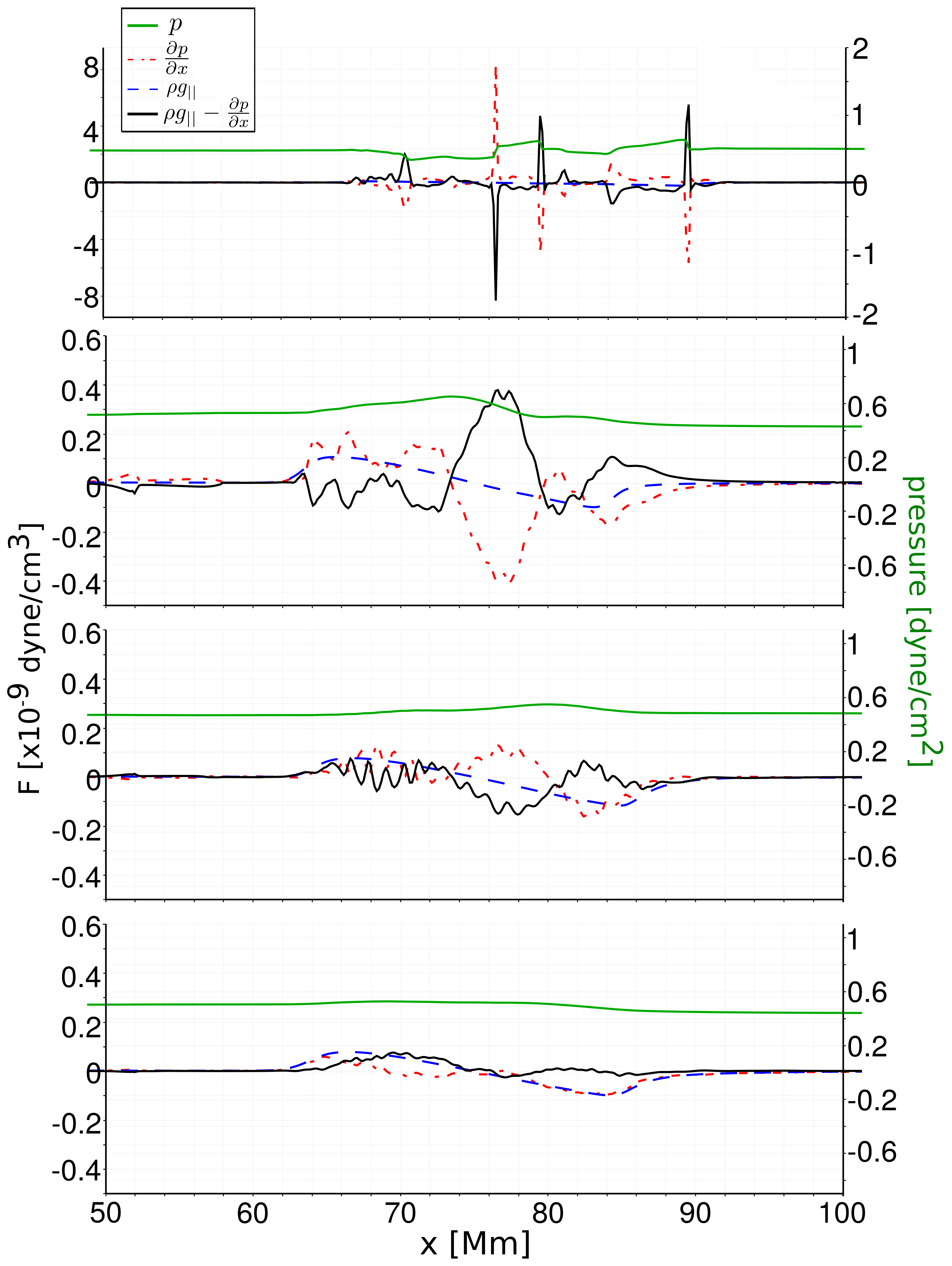}
    \caption{Plot of the $x$ component of forces acting on thread \textit{4}. The dash-dotted red line denotes the pressure gradient component $\frac{\partial p}{\partial x}$, the dashed blue line denotes the force due to gravity $\rho g_{||}$ and the full black line represents the sum of forces along $x$ ($\rho g_{||}-\frac{\partial p}{\partial x})$. The green line is gas pressure.}
    \label{fig:forces_x}
\end{figure}

   Figure~\ref{fig:forces_x} shows the forces acting on one of the threads (in this case thread \textit{4}, but the same is applicable for others). The full black line corresponds to the total force. We plotted the pressure gradient (dash-dotted red line) with the positive sign, however, accounting for the minus sign it is the one predominantly responsible for the total force we see. The top panel shows the forces around the time the threads reach their maximum deviation, at $t=30$\,min (corresponding to the first panel of Fig.~\ref{fig:snapshots}). The pressure gradient force reveals compressional, slow wave modes propagating through each thread. We recognise them as slow modes, as their time to cross the thread corresponds to the sound speed, that is, to the velocity of the slow magnetoacoustic waves. They are particularly distinctive earlier on in the oscillation. The same is also noticeable on the $y$-scale. The values describing the forces are an order of magnitude larger on the top panel than on the other three bottom panels. The more time progresses, the more those compressional waves get attenuated. As the various waves are repeatedly encountering PCTR and TR-left and TR-right interfaces where reflection and transmission occurs, they continuously constructively and destructively interfere. The interference happens with new waves entering the thread and with the existing waves inside the prominence reflected at the PCTR. All of that changes their initial sharp appearance. Half a period later (at $t=71$\,min), when the threads reach their maximum deviation away from their equilibrium position, those compressional waves are less prominent but still surpass the value of the gravity force. Over time, they continue diminishing and it is already at $t=138$\,min that we can  see that the system is again close to equilibrium.

\begin{figure}
    \centering
    \includegraphics[width=0.45\textwidth]{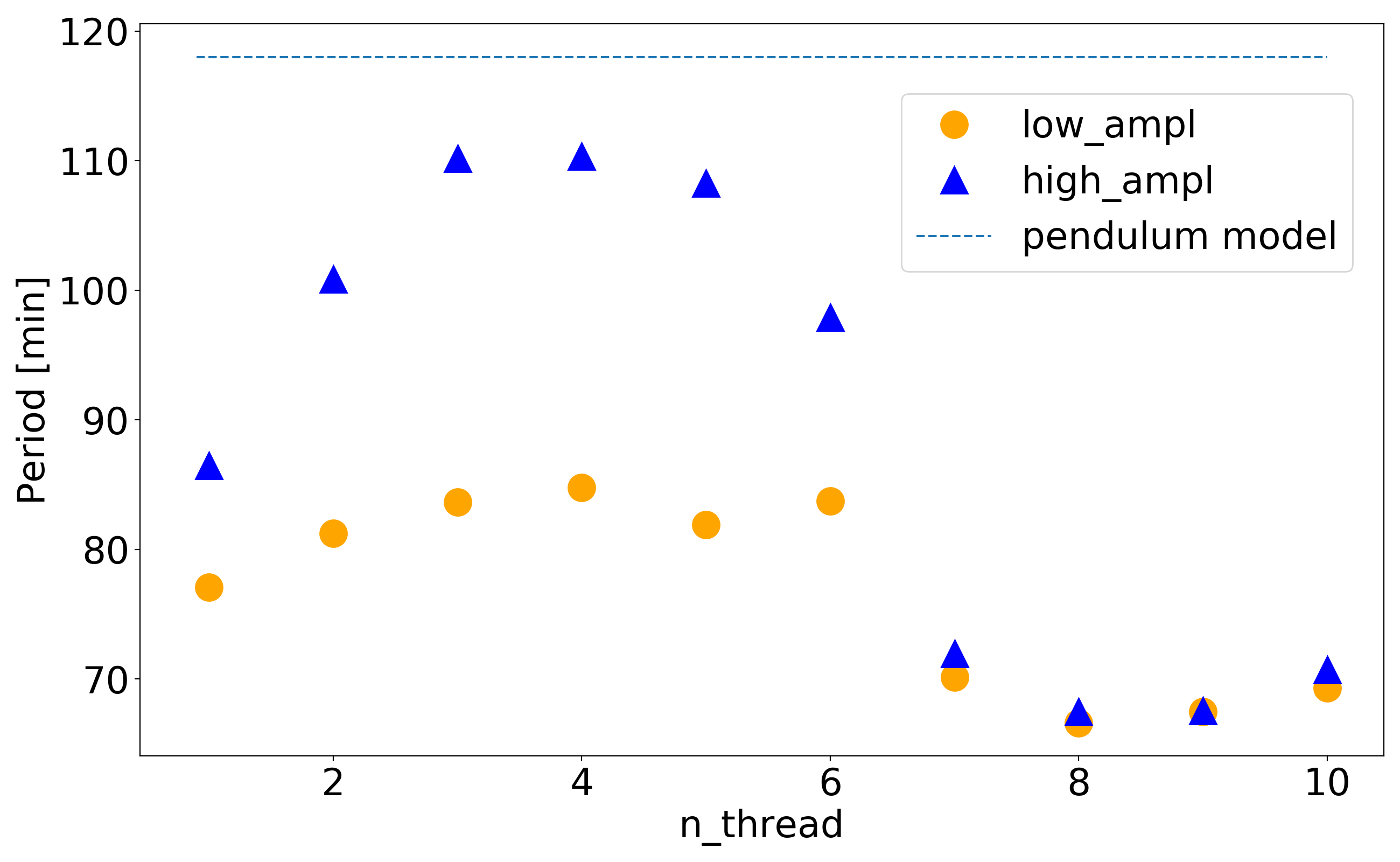}
    \caption{Values of period calculated by fitting function $f(t)$ (Eq.~(\ref{eq:fitting_func})) to $d_x^{CM}$, for each thread (\textit{1}--\textit{10}). The circles represent values for low\_ampl while the triangles represent values for high\_ampl and the dashed horizontal line denotes value of period that we expect from the pendulum model.}
    \label{fig:period_comparison}
\end{figure}

   The only forces that are able to play the role of a restoring force in this scenario are those that are due to the gravity component and the pressure gradient force. The influence of the source on the measured period is stronger for threads in the direct path of the shock wave (\textit{1}--\textit{6}) and when the amplitude of the source is larger. If we inspect and compare the displacement periods of the threads in the low\_ampl and the high\_ampl case (Fig.~\ref{fig:period_comparison}), we can notice that in the high\_ampl case, the periods are actually longer. Also, there is no great difference in the periods for threads \textit{7}--\textit{10}, which were not directly hit by the shock wave. In the catalogue of prominence oscillation, \cite{Luna2018} describe two particular prominences where the same prominence exhibits two very different oscillations (all four events have oscillations with velocity amplitudes $<$10\,km\,s$^{-1}$, same as in our results). The observed oscillations differ in period and damping time: in the first prominence, the second oscillation even shows amplification. These authors did not make any definitive conclusions on what exactly caused the change of the period. They did note, however, that the first prominence that exhibited this peculiarity was surrounded by nearby flares; thus, they stated that it is quite plausible that the influence of an external driver could be responsible for the different periods. Besides these two cases where \cite{Luna2018} gave a more detailed description, there are more cases in their catalogue that show the same characteristic, a single prominence with different periods of oscillation. They concluded that such events are most likely related to the nearby external drivers. \cite{Zhang2019} already described how in the adiabatic case the pressure gradient acts as a restoring force. Additionally, in this case, it is also necessary to take into account the contribution of the constructive and destructive interference of the propagating waves (initially resulting from the source region) to the restoring force. Accordingly, we notice a clear deviation of our measured period from the simple monolithic pendulum model and, thus, we see that in order to more precisely estimate the period, additional factors need to be taken into account. An important role in the resulting period of prominence oscillation comes from the influence of the driver, as well as the fact that a multi-threaded structure necessarily gets perturbed with distinct delays between individual threads, inducing interference. 

   Furthermore, since the pendulum model (dashed line in Fig.~\ref{fig:period_comparison}) is only a first approximation when estimating the correct value of the period, other factors are to be considered. \cite{G&T2020} showed that the bigger the radius of the curvature in which the prominence resides, the more significant differences are found with the pendulum model. The reason is a larger contribution of the pressure gradient as the restoring force. In our case, both the low\_ampl and the high\_ampl cases are done in the same curvature setup with the radius of the curvature of the arcade field lines being $\sim$\,331\,Mm. Another point that \cite{G&T2020} clearly showed, is that the period, also depends significantly on the width of the prominence  (besides
the geometry of the coronal arcade and the pressure gradient). According to their results, the wider the prominence, the longer the period. Considering that our case has a very elongated domain in comparison to its width, we do expect this to influence the measured period with respect to the analytical model.

   From previous 1D simulations, having a similar dipped magnetic field line topology \citep{Xia2011,Zhang2012,Zhang2013,Zhou2014,Zhang2020} and the fact that the motion happens dominantly along the magnetic field lines, we do not expect significant changes in the magnetic field evolution. Even so, the magnetic field oscillates with low-frequency oscillations in relation to the changes in density and thermal pressure (Fig.~\ref{fig:CMx_Btot_mass_thr4} and \ref{fig:p_t_rho_thr4}). Thanks to the ideal MHD, frozen-in approximation, the plasma motion takes into account the magnetic field lines and perpendicular motions displace the field lines. Nonetheless, there is still a need to explain the oscillations seen in the magnetic field, as well as the pressure and density that can be clearly seen in Figs.~\ref{fig:CMx_Btot_mass_thr4} and \ref{fig:p_t_rho_thr4}. From the Fourier analysis, we get a period of 27.4 min, which does not correspond to the period of longitudinal oscillations (cf. Table~\ref{table:displ}). The period of 27.4\,min does not appear to be relatable to the strength of the flare, since nearly the same period is measured in high\_ampl case (28.2\,min). In the work done by \cite{J&R1992a}, the authors theoretically analyzed modes of oscillation in a quiescent prominence where the magnetic field is aligned with the prominence sheet (without gravity). The existence of the fundamental slow body modes that they show is of particular interest in the present context as well. Taking into account the dimensions of our system and the approximate formula for deriving the period \citep[cf. Table. 1 in][]{J&R1992a}, we get a value of the period of $\sim$29\,min, which closely corresponds to the measured periods of oscillations seen in pressure, density, temperature, and the magnetic field. This means that the oscillations in the mentioned parameters correspond to the threads oscillating with a fundamental slow body mode. 

   Besides the described slow wave modes seen (nearly with no difference) in the low\_ampl and high\_ampl cases, the high\_ampl case generally shows the same pattern of motion as the low\_ampl. Because the source has the same position and radius, the global motion of threads is equivalent. As a result of the source having double the amplitude, the changes in each parameter are greater, especially for threads that were hit directly by the shock (\textit{1}--\textit{6}). The compressions and rarefactions in the plasma are more intense and, as a consequence, we see more significant changes in the magnetic field components in the high\_ampl in respect to the low\_ampl case. More mass of the threads gets redistributed to coronal matter (due to the contour-based identification). Along with it, we measure higher average temperatures of the threads, attributable to the stronger compression of the plasma resulting from the larger amplitude shock.

\subsection{Transverse oscillations}

   As a result of the initial transverse movement of threads in the $y$ direction (Fig.~\ref{fig:dy}), a pressure gradient results and, subsequently, a Lorentz force appears. The sum of these two forces in the $y$ direction causes oscillatory motion of the threads. The resulting changes in the pressure and density along the $y$ direction induce changes in the $y$ component of the magnetic field, similar to the $x$ direction. The distribution of the initial transverse movement of the threads (seen in Fig.~\ref{fig:dy}), is determined by the relative position of the source and the distribution of threads with regard to the source. \cite{Zhang2019} showed that in a 2D adiabatic simulation, such transverse oscillations (vertical in their case) resulted in the leakage of energy in the form of fast waves. They identified that as one of the main reasons of damping. However, as a consequence of the periodic boundary conditions along our transverse $y$ coordinate, a standing wave is formed across the direction of the threads. In other words, with our setup, the wave leakage doesn't result in a loss of energy but solely in its transmission from one to the other side of the domain along the $y$ direction. Since the change of $v_y$ and $B_y$ are significantly smaller for threads \textit{4} and \textit{9} than for any other thread, it is reasonable to assume that the $y$ coordinates of those threads (-900\,km and 2100\,km, respectively) correspond to the position of the nodes of the standing wave (wavelength of 3\,Mm). The changes in the parameters of threads \textit{4} and \textit{9} are not exactly zero, hence, there is a transfer of energy across the direction of the field lines. This is the most probable reason for the amplification of the longitudinal motion seen in threads \textit{8}, \textit{9,} and \textit{10}.

   To explain the numerous high-frequency oscillations clearly evident in all the parameters, we made an estimate of the velocity of the fast and slow waves and their crossing times (because of the low-beta regime, the Alfv{\'{e}}n waves have approximately the same value of velocity as fast waves). The crossing times in the direction perpendicular to the threads, for either of the MHD waves, is on the order of a few seconds and does not seem to cause any of the observed high-frequency oscillations. As for the parallel propagation, the time it takes for the slow waves to pass through the coronal area, between the TR and the edge of prominence, is about 3\,min; for the fast waves it is about 1\,min. Additionally, we can also consider the wave propagation through the prominence plasma. The slow (sound) velocity, internal to the threads is around 26\,km\,s$^{-1}$ and of the fast wave is about 78\,km\,s$^{-1}$. In that case, for slow waves we get a crossing time through a thread of a length of about 21\,Mm to be around 13.5\,min, for fast waves the crossing time is about 4.5\,min. Considering that the length of each thread changes as the threads oscillate (starting with the value of 23\,Mm after the relaxation phase and dropping to values below 19\,Mm), so do the crossing times of each wave. Since the high-frequency oscillations all primarily take place in the range of 1.5 to 16\,min, we argue for the role of magnetoacoustic waves propagating through the threads and the corona, as the cause of such high-frequency oscillations. 

\subsection{Energy analysis}
\label{sec:energy_analysis}

   To give a complete description of the motion taking place in this scenario, we can also describe it from the energy perspective. In modelling the source region (as described in Sect.~\ref{sec:method}) we introduced additional energy into the system. When the resulting shock wave hits the threads, it transfers part of its associated energy to them and sets them into oscillatory motion. As shown in other works \citep{Luna2016,G&T2020}, the main, longitudinal oscillations show strong damping (Table~\ref{table:displ}), especially in contrast to the transverse horizontal oscillations, which remain unattenuated for the entire simulated time (Fig.~\ref{fig:dy}). \cite{Luna2016} speculated that the stress created by numerical viscosity, could produce significant damping. They experimented with the influence of such stress with grid sizes of 500 and 250\,km. Our resolution is significantly finer (36$\times$7.5\,km) than their highest resolution, thus ensuring that is not an issue here. In addition, \cite{Liakh2021} showed that with the dimension of the grid cell of about 30\,km, it is sufficient to avoid numerical damping and it is possible to investigate the real physics causes. Moreover, our conservative shock-capturing method does not rely on any added artificial viscosity (but has numerical dissipation which is on a very low order as this scales with the grid size). Then, another possible reason for damping is wave leakage, demonstrated in the work by \cite{Zhang2019}. The authors found fast waves propagating in the vertical direction with respect to the magnetic field. In our setup, the waves transverse to the field get recycled by the periodic boundary assumption and do not allow for the escape of energy (waves). To further analyze the damping, we performed a 1D simulation of oscillation of a single thread \citep{Jercic2021}. The magnetic topology and the boundary conditions are exactly the same. Because we are working with an ideal MHD, such a setup necessarily becomes a pure hydrodynamic, adiabatic evolution. Once we introduce the mass into the system, there is no way for the oscillations to dampen. As we bring the source into the 1D domain, prominence oscillations are induced by the shock wave, in the same way as in the 2D simulations. However, the oscillations in this 1D case show no damping whatsoever. Hence, a similar purely 1D hydro representation of a single thread does not show pronounced damping. This implies a role for the 2D nature adopted in the main simulation analyzed here, with the exchange between longitudinal and transverse wave energy leading to the damping of the individual longitudinal thread motions.

\begin{figure}
    \centering
    \includegraphics[width=0.45\textwidth]{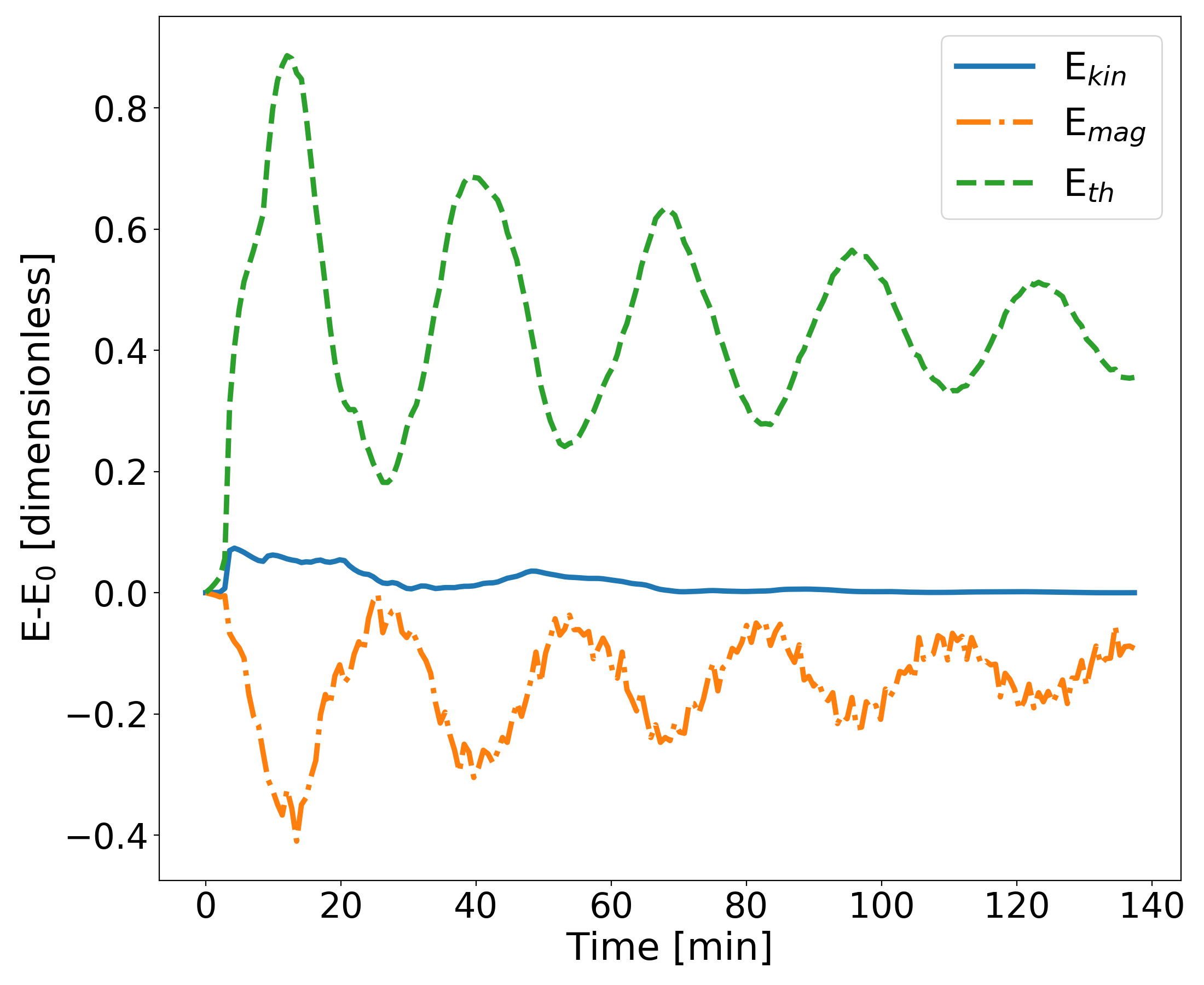}
    \caption{Changes of the average thermal (green dashed line), kinetic (full blue line), and magnetic (dashed-dotted orange line) energy of thread \textit{4} with respect to their initial value (after the relaxation phase).}
    \label{fig:energy_thread4}
\end{figure}

   In order to illustrate the energy changes in our 2D system, we calculated the average components of the total energy of each thread:

\begin{equation}
    E_{kin}(t)= \frac{1}{2}\frac{\int_{\mathcal{A}} \big[v_x^2(t,\Vec{x}) + v_y^2(t,\Vec{x})\big]\rho(t,\Vec{x}) dA}{\int_{\mathcal{A}} dA} \,,
\end{equation}

\begin{equation}
    E_{mag}(t)= \frac{\int_{\mathcal{A}} \big[B_x^2(t,\Vec{x}) + B_y^2(t,\Vec{x})\big] dA}{\int_{\mathcal{A}} dA} \,,
\end{equation}

\begin{equation}
    E_{th}(t)= \frac{1}{\gamma -1}\frac{\int_{\mathcal{A}} p(t,\Vec{x}) dA}{\int_{\mathcal{A}} dA} \,,
\end{equation}

   \noindent as shown in Fig.~\ref{fig:energy_thread4}. Several elements contribute to the damping seen in 2D and not in 1D. First, in 2D, part of the energy the threads receive from the shock wave is converted to thermal energy (an increase in temperature); whereas in 1D, we don’t see that same effect. In 1D, the temperature actually drops instead of increases. We speculate that the triggering wave that travels unhindered in between the various threads thereby compresses the threads laterally, an effect that is completely overlooked in a 1D study. The plot of the thermal energy in Fig.~\ref{fig:energy_thread4} (green dashed line) shows that the thermal energy at the end of oscillation is obviously larger than the value at the beginning. Secondly, because the thread is identified through the density threshold, we can notice in Fig.~\ref{fig:snapshots} that this does not fully represent all plasma that initially resided in the initial thread, as the thermodynamic changes are seen on a larger area. The plasma of the threads is lost and mixed with the coronal plasma (which also explains why the threads lose mass; see Fig.~\ref{fig:CMx_Btot_mass_thr4}). In that way, the associated energy of the shock wave is converted into the mechanical and thermal energy of the threads and the thermal energy of the plasma that escapes (now part of the surrounding coronal plasma). Thirdly, even though we do not have any loss of energy from the system resulting from the wave leakage as in \cite{Zhang2019}, we still have a similar effect taking place. Because of the transverse oscillations (inherently not possible in 1D) there is an exchange of energy from the threads directly hit by the shock to the side threads, more specifically, to threads \textit{8}--\textit{10} (Fig.~\ref{fig:CMx8_10}). As for the influence of the source itself on the damping, by comparison of low\_ampl and high\_ampl cases, we noticed that the damping is increased in the high\_ampl case (lower $\tau$/P ratio). That can be explained by the correlation between attenuation and oscillation amplitude \citep{G&T2020}. 
   
   The changes seen in the magnetic energy (orange dashed-dotted line) in Fig.~\ref{fig:energy_thread4} can be explained by the relation of the magnetic field to the motion of plasma due to the frozen-field assumption. From Fig.~\ref{fig:energy_thread4}, we see that the kinetic energy (solid blue line) decays quickly, with the first two oscillations most prominent, after which it diminishes almost completely. The energy carried by the shock wave is distributed among all the threads. Most of it is transferred to thread \textit{4}. As part of the total energy, the kinetic energy has the lowest contribution, nonetheless, it is still the one responsible for all the motion that we see in the system.

\begin{figure}
    \centering
    \includegraphics[width=0.45\textwidth]{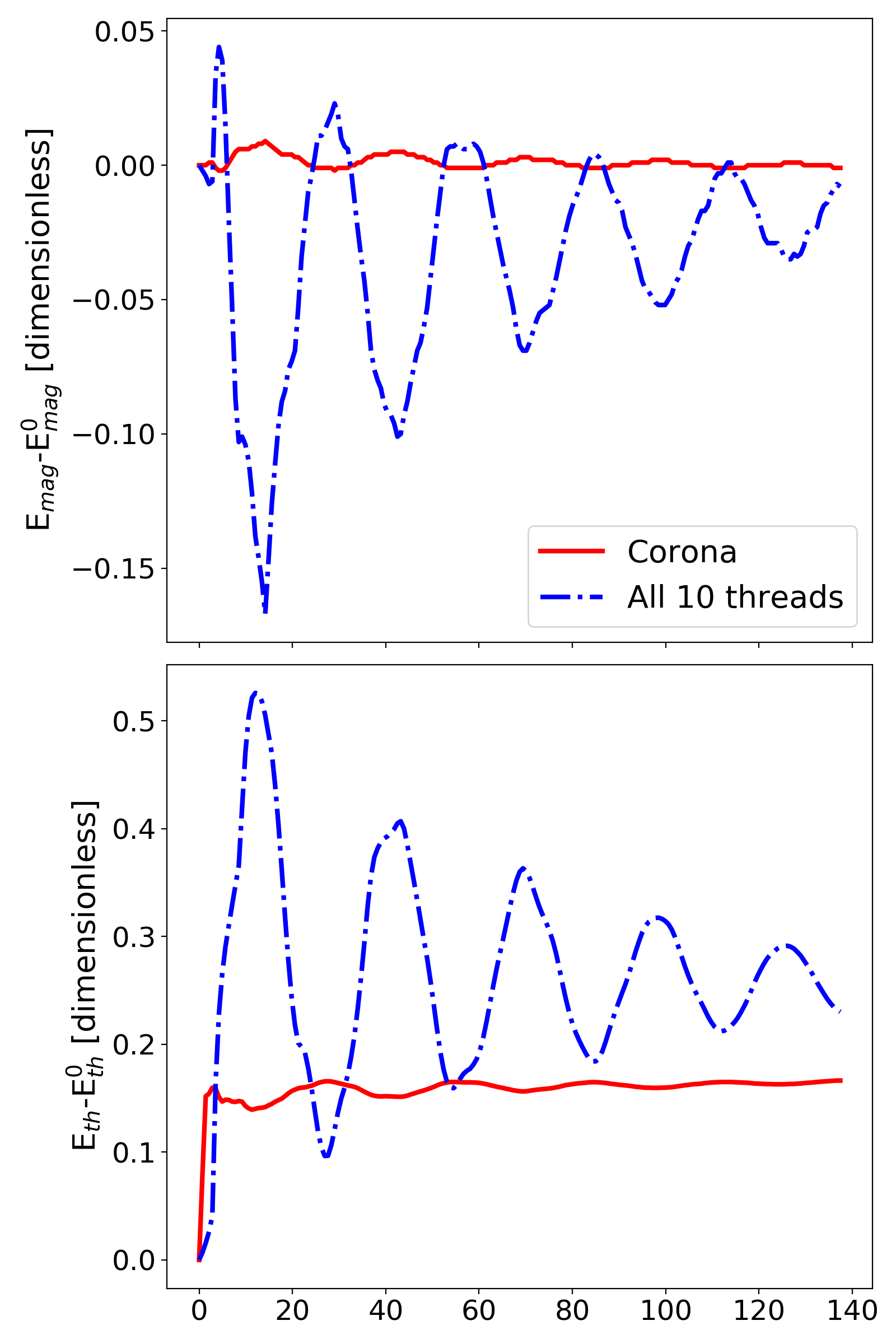}
    \caption{Changes of the average magnetic (top panel) and thermal (bottom panel) energies of all ten threads and the corona, with respect to their initial value (after the relaxation phase).}
    \label{fig:energy_corona_threads}
\end{figure}

   We additionally explored the interchange of energy between the ten threads and the corona (with number density $<$5$\times$10$^{-10}$\,cm$^{-3}$), which the threads are in direct contact with. From Fig.~\ref{fig:energy_corona_threads}, we see that the changes of the area-averaged energies are opposite in the coronal and prominence plasma. In particular, as one increases, the other decreases, further demonstrating the important interplay between the two related systems.

\section{Conclusion} \label{sec:conclusion}

   The goal of this paper is to study the detailed response of a realistic multi-threaded prominence to the passing of a coronal shock wave. This allows us to address the as yet unknown role of coupled longitudinal and transverse oscillatory motions and to carry out an analysis of how the individual threads behave with a realistic source of perturbation imposed on the domain. We showed that the induced motion is dominantly longitudinal. However, its period shows discrepancies with the values expected from the simple monolithic, analytical pendulum model. There are several factors that we identified as the cause of the deviation. First and foremost, there is the influence of the source region. The shock wave that propagates through the domain creates compressional waves propagating through the thread that reflect off of the PCTR and interfere with each other. These waves also travel through the coronal plasma in between the threads, thereby causing additonal compression and providing a contribution to the restoring force. The nature of the contribution depends on whether they interfere constructively or destructively at the moment when the threads are returning or moving away from their equilibrium position. The fact that the radius of the curvature of the arcade field lines is large and the treads are relatively narrow in comparison to the length of the domain add to the deviation of the period value from the expected pendulum model. We note that this deviation is relevant for prominence seismology, since by measuring the period from observational data we can estimate the coronal magnetic field.

   Increasing the strength of the source and keeping its position and size the same, the global motion of the threads does not change. However, it clearly induces a greater change of the displacement and velocity amplitude, while also affecting the thread pressure, density, magnetic field, and temperature. The period additionally lengthens and the damping intensifies.

   Besides the dominant longitudinal motion, transverse oscillations are also present. They get increasingly pronounced with a larger amplitude of the source of the shock wave. We detect an exchange of momentum in the direction transverse to the magnetic field. It results in the amplification of longitudinal motion of threads \textit{8--10}. High-frequency oscillation are detected in $d_y$, $v_y$, $B_x$, and $B_y$ in low\_ampl and high\_ampl cases, with periods in the range of 1.5 to 16\,min. Calculating the crossing times of magnetoacoustic waves in the corona between the TR and the threads (PCTR) and also along the length of the threads themselves, we notice that all the crossing times appear in the range of observed high-frequency oscillations (1.5 to 16\,min). From this analysis, we can already comment on the necessary resolution of instruments to observe the transverse displacements identified here. We  mention above that the amplitude of the longitudinal motion increased by a factor of two, following the increase in the amplitude of the source. A similar relation is noticeable for the transverse oscillation. These get significantly more pronounced with a stronger flare (Fig.~\ref{fig:dy}). Keeping in mind that the strength of the source that we implemented coincides with that of a microflare, we can expect larger scales of transverse oscillation on the Sun. The resolution of the Atmospheric Imaging Assembly onboard the Solar Dynamic Observatory spacecraft is 1.5\,arcsec ($\sim$1088\,km) on its 1\,AU distance from the Sun. The Wide-Field Imager for Solar Probe Plus on board Parker Solar Probe and the Solar Orbiter Heliospheric Imager on board Solar Orbiter are expected to have even better resolutions considering they will reach approximately 0.3\,AU from the Sun. Using images from the new spacecraft, we just might be able to capture the oscillations that we are already able to simulate.

   Lastly, by comparing our 2D simulation with its counterpart in 1D, we discuss the probable reasons for the damping seen in our 2D simulation and yet completely absent in 1D. Considering the high resolution we employ, we can exclude the role of numerical viscosity on damping \citep[cf. ][]{Liakh2021}. Prevailing causes of damping include: the conversion of the initial energy transferred from the shock wave to the threads and the resulting redistribution of the prominence mass to the corona. Another characteristic of the 2D simulation is that transfer of energy is now possible in the $y$ direction. That fact explains the transfer of energy from threads \textit{1}--\textit{6} (hence, the damping of their longitudinal motion) to threads \textit{8--10} (hence, the amplification of their longitudinal motion). If we consider the possibility of having a third direction, we can expect the oscillation to become further enriched. We can already see how  going from 1D to 2D changes the results of numerical simulations considerably. An additional degree of freedom has influenced the observed damping in particular. We can expect a third direction to additionally add to the possibility of transverse oscillations in the vertical plane and transferring energy in that direction \cite[as was already shown for the vertical plane by][]{Liakh2021}.
   
   In the case of more in-depth parametric follow-up studies, the arcade geometry could be changed, as well as improvements to the model (e.g. including non-adiabatic effects). This would allow us to gain a deeper understanding of the interplay between the source properties and the resulting oscillation. This can also be used to study the thread formation in ab initio models \citep[as in][]{Zhou2020}, which are then perturbed by external shock waves. These studies would then have a direct impact on the interpretation of observations and on our understanding of as-yet-unexplained prominence dynamics.

\begin{acknowledgements}
      VJ acknowledges the funding from Internal Funds KU Leuven. Authors are supported by the ERC Advanced Grant PROMINENT and a joint FWO-NSFC grant G0E9619N. This project has received funding from the European Research Council (ERC) under the European Union’s Horizon 2020 research and innovation programme (grant agreement No. 833251 PROMINENT ERC-ADG 2018). Further, this research is supported by Internal funds KU Leuven, project C14/19/089 TRACESpace and FWO project G0B4521N. The visualisations were achieved using the open source software \href{https://www.paraview.org/}{ParaView}, \href{https://www.python.org/}{Python} and \href{https://yt-project.org/}{yt}. The resources and services used in this work were provided by the VSC (Flemish Supercomputer Center), funded by the Research Foundation - Flanders (FWO) and the Flemish Government. VJ would also like to thank Joris Hermans for always fun and fruitful discussions.
\end{acknowledgements}

%
%

\bibliography{42127}{}
\bibliographystyle{aa}

\end{document}